\documentclass[pre]{article}
\usepackage[latin9]{inputenc}
\usepackage{float}
\usepackage{amsthm}
\usepackage{amsmath}
\usepackage{amssymb}
\usepackage{graphicx}
\usepackage{hyperref}

\makeatletter

\numberwithin{equation}{section}
\numberwithin{figure}{section}

\usepackage{bbm}\usepackage{psfrag}

\makeatother

\begin{document}

\title{Behind the price: on the role of agent's reflexivity in financial market microstructure\footnote{Authors acknowledge partial support by the grant SNS13LILLB ''Systemic risk in financial markets across time scales". }}
\author{Paolo Barucca$^1$ and Fabrizio Lillo$^{2}$\\
\textit{\small1. Department of Banking and Finance, University of Zurich, Switzerland}\\
\textit{\small2. Scuola Normale Superiore di Pisa, Pisa, Italy}}
\maketitle
\begin{abstract}
In this chapter we review some recent results on the dynamics of price formation in financial markets and its relations with the efficient market hypothesis. Specifically, we present the limit order book mechanism for markets and we introduce the concepts of market impact and order flow, presenting their recently discovered empirical properties and discussing some possible interpretation in terms of agent's strategies. 
Our analysis confirms that quantitative analysis of data is crucial to validate qualitative hypothesis on investors' behavior in the regulated environment of order placement and to connect these micro-structural behaviors to the properties of the collective dynamics of the system as a whole, such for instance market efficiency. 
Finally we discuss the relation between some of the described properties and the theory of reflexivity proposing that in the process of price formation positive and negative feedback loops between the cognitive and manipulative function of agents are present. 

\end{abstract}
\pagebreak{}

\section*{Introduction}

Understanding price movements, both their origin and their properties, is one of the most important challenges in economics. In Finance this problem has a long history which has seen two antagonist schools of thought confronting in the first half of the twentieth century. On one side there were the fundamentalists who posited that the price of an asset is the discounted value of its "intrinsic" or "fundamental" value and equals the discounted cash flow which that security gives title to. On the other side there were the econometricians, who, applying statistical analyses to empirical time series of prices discovered that stock prices develop patterns that look like those of a random walk. The latter is the time series that can be obtained, for example, by tossing a coin and moving up (down) the price by one unit when the outcome is head (tail). The erratic and random behavior of prices seemed to clash with the fundamentalist view and seemed to give support to those believing that stock market is essentially a casino. By using the words of LeRoy "If stock prices were patternless, was there any point to fundamental analysis?". It is well known that the solution to this problem was given by the seminal 1965 paper by Paul Samuelson  "Proof that Properly Anticipated Prices Fluctuate Randomly" (even if Bachelier 1900 and Cowles 1933 arrived to somewhat similar conclusions). Samuelson showed that in an informationally efficient market, i.e. a market which fully incorporates the expectations and information of all market participants, price changes must be unforecastable. This is the celebrated Efficient Market Hypothesis (EMH), a cornerstone of modern Finance, which reconciles the fundamentalist and econometrician view. 

However the details of how information is impounded into price are still a matter of debate, as well as the question of whether markets are truly efficient. Market microstructure is devoted to the empirical and theoretical study of how information (private or public) is incorporated into price and how price is formed through the action of many interacting agents. Given the importance of the problem (even outside Finance, think for example to the new markets for advertisement on search engines in the Internet), the increasing availability of high resolution data, and the significant changes observed in the organization of markets, market microstructure has experienced a large development in the last fifteen years. In this paper we review the problem of price formation from
a microstructure point of view focusing on how strategies are slowly translated into orders in the market. We do not aim to be exhaustive, but to highlight the main elements which have recently emerged in the field, trying to avoid as much as possible technicalities and the use of mathematical formalism.  In the last part of the paper we will discuss about possible analogies and relations between the newly discovered properties of price formation and the theory of social reflexivity, proposed, among others, by George Soros. 

The contribution is divided in four sections. In the first section
we present the efficient market hypothesis, the concept of market impact, and we describe a widespread market mechanism, namely the limit order book. In the second section we present the order flow and describe one important properties, its long memory, which allows us to understand the strategy of order splitting followed by large investors. In this section we also discuss how the long memory of order flow can be reconciled with efficient market hypothesis. In the third section we review Soros' theory of social reflexivity and finally in the last section we outline our
interpretation of empirical results in the context of social reflexivity.

\bigskip{}

\section{Market impact and the limit order book}

The market is where supply and demand meet, that is where buyers and
sellers exchange goods. The efficient-market hypothesis
(EMH) states \cite{key-23,key-24} that stock market efficiency implies
existing share prices to always incorporate and reflect all relevant
information. According to EMH, any information in the hand of an investor on the future value of an assets should thus be instantaneously incorporated into the price. This occurs through the choice of the strategy the investor chooses to trade the asset. The trading strategy can be seen as the medium that translate information into price changes. 
Loosely speaking {\it market impact} refers to the correlation between an incoming order or strategy (to buy or to sell) and the subsequent price change. The strategy can be coded in the series of {\it orders} to buy or to sell, which are sent to the market. The {\it order flow} is the aggregation of the orders of all the agents (or a subset of them).

However, the causal relation between trading strategies (or order flow) and price changes is far from trivial and at least three explanations for this correlation (i.e. the existence of market impact) can be given
\begin{itemize}
\item Agents successfully forecast short term price movements and trade accordingly. This does result in measurable correlation between trades and price changes, even if the trades by themselves have absolutely no effect on prices at all. If an agent correctly forecasts price movements and if the price is about to rise, the agent is likely to buy in anticipation of it. According to this explanation, trades with no information content have no price impact.
\item The impact of trades reveals some private information. The arrival of new private information causes trades, which cause other agents to update their valuations, leading to a price change. But if trades are anonymous and there is no easy way to distinguish informed traders from non informed traders, then all trades must impact the price since other agents believe that at least of fraction of these trades contains some private information, but cannot decide which ones.
\item Impact is a purely statistical effect. Imagine for example a completely random order flow process, that leads to a certain order book dynamics. Conditional to an extra buy order, the price will on average move up if everything else is kept constant. Fluctuations in supply and demand may be completely random, unrelated to information, but a well defined notion of price impact still emerges. In this case impact is a completely mechanical - or better, statistical - phenomenon.
\end{itemize}

\begin{figure}\label{lob}
\centering
\includegraphics[scale=0.5]{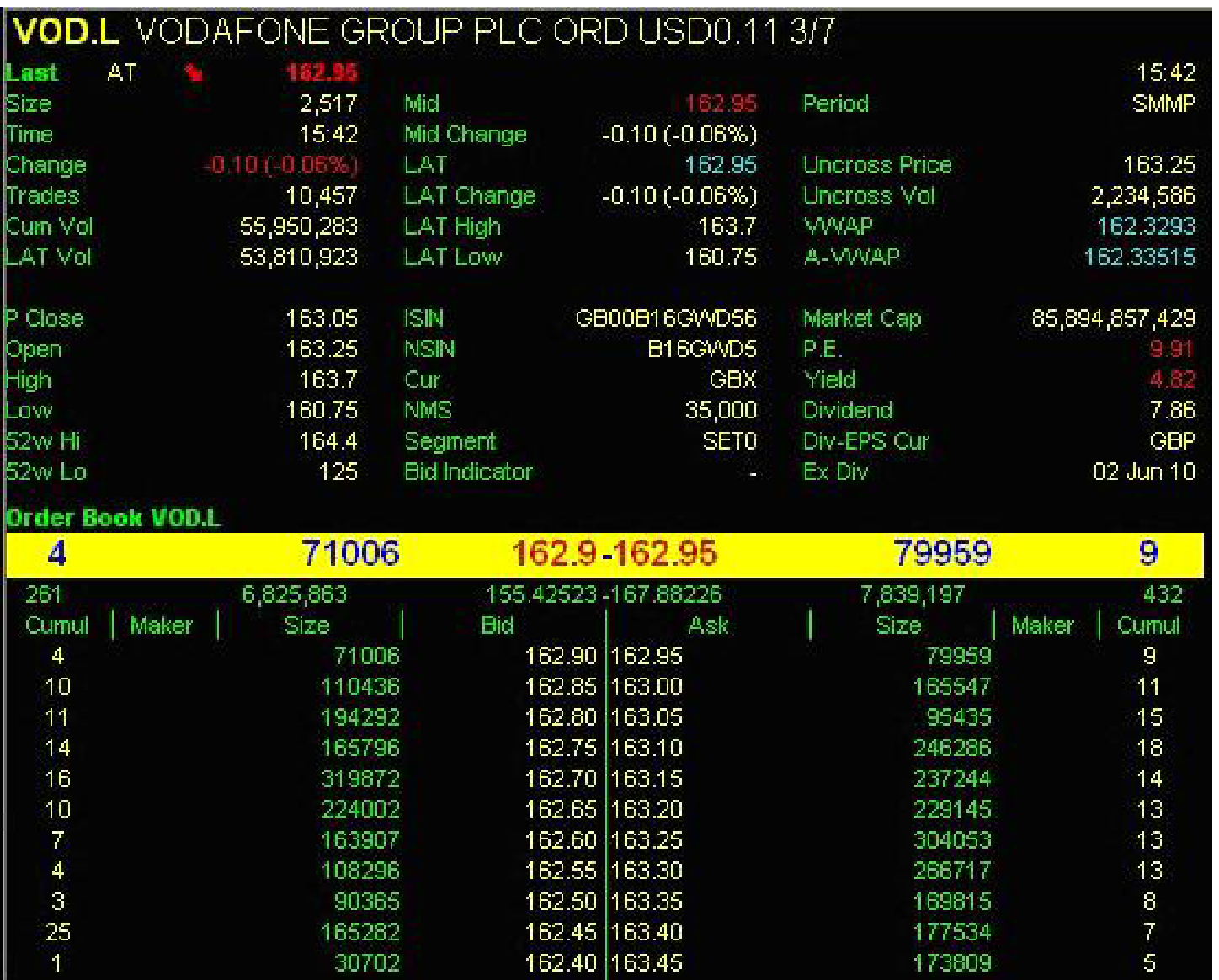}
\protect\caption{The typical structure of an order book. \cite{key-27}
}
\end{figure}

The distinction between these three scenarios is not fully understood. In order to discriminate among these alternatives, it is useful to specialize into a concrete case, since markets have different structures and rules. In other words, one way to address this problem in a quantitative fashion is through
dynamical models of market microstructure, which are based on a specific,
'internalist', knowledge of what kind of orders can be placed by investors
and what is their dynamics. Here we will focus on the {\it limit order book}, a very common mechanism adopted in most electronic markets. In a limit order book market, an agent (or an intermediary acting on her behalf) can place two types of orders\footnote{Even though various financial markets may have more kinds of slightly
different orders, these are the two main types-}, namely limit and market orders:
\begin{itemize}
\item Limit orders are orders to buy or sell a given
volume of shares at a specified price or better. 
\item Market orders are orders placed to buy or sell an investment immediately
at the best available current price.
\end{itemize}
The best buy and sell order on the market are called ask and bid respectively
and their difference is the bid-ask spread. Buy (sell) limit orders with a price lower (higher) than the ask (bid) do not initiate a transaction and are stored, visible to other market participants, waiting for the price moving in their direction. An agent can decide to cancel a limit order at any time. Figure \ref{lob} shows a snapshot of a real order book.

Orders are not typically placed directly by individual investors, mutual funds
or hedge funds: all these economic agents need to ask intermediaries
to place orders of given volumes of various stocks on their behalf.
Other subjects that can place orders in the market are market makers
who act as intermediaries for the market as a whole: they continuously
place orders in the order book so to provide 'liquidity' to the market,
that is to give investors the chance to buy and sell anytime at their
will. Market makers take risks by buying and selling countercurrent
but make profit through the bid-ask spread.

Market impact arises for two reasons. The first is purely mechanical. A market order with a volume larger than or equal the volume at the opposite best (the bid for a sell market order and the ask for a buy market order) will move mechanically the price, increasing the spread and therefore the mid price\footnote{Note also that if market order volume
is smaller than or equal to the volume at the opposite best,  the order is executed at the best price; on the other hand if its volume exceeds
the volume of the opposite best in the order book, then
it penetrates the book and reaches the second best price or more.
In this case the price is a weighted average over the various limit
orders that are needed to execute the market order.}. Notice that even an uninformed trade moves mechanically the price and there is no relation between information of the trade and price movement (see the third explanation in the bulleted list above). This is even reinforced by the fact that electronic markets are typically anonymous and there is no way of knowing the identity of the counterpart. 

The second possible origin of market impact is due to the reaction of the rest of the market to the trade. Even if the market order does not move mechanically the price, the other agents might react to the observation of the trade, revising the price of their limit orders and thus moving the price. As we will see below, the induced market impact plays an important role. In general, the more liquid is the market, i.e. the
more limit orders are present in the order book then the lower is
the market impact and the less a single investor can induce large price
changes. 

A number of empirical studies \cite{key-13,key-28,key-14} has established that market impact, both mechanical and induced, is statistically different from zero (despite being quite noisy). This means that buy (sell) market orders move on average the price upward (downward). Generically it is found that the larger the volume of the trade the larger on average the price change and the dependence between these two quantities is strongly concave. In simple words this means that if the volume of a market order is doubled, the price impact is significantly less than doubled. The fact that even uninformed trades can change the price (again, also because it is almost impossible that the market understands quickly if the trader is informed or not) raises several interesting questions on the relation between information and price changes and therefore on the Efficient Market Hypothesis. Anticipating the discussion we will do later in this paper, it suggests the presence of positive feedback loops, where an uninformed trade transiently impacts price, but the other market participants, being unable to discern if the trade is informed or not, revise their valuation of the asset and trade concurrently, creating more price change in the direction of the random initiator trade. This amplification mechanism resembles the reflexivity hypothesis (see below for more details). 

Note that all the above analysis is static, i.e. it refers to the contemporaneous relation between trades and price changes. An important aspect to elucidate better the process with which markets understand if a trade is informed or not and how the information (or the lack thereof) is incorporated into price requires an inter-temporal analysis, investigating how the orders arrive to the market sequentially in time and how they impact the price. Next Section discusses some recent findings in this direction. 

\section{The long memory of the market}

In the last ten years there have been major efforts to understand
investors behaviors through detailed modeling and data-analysis. A significant amount of the 
literature has been devoted to analyze data on order books of real
financial markets, \cite{key-11, key-13}. In particular we review here the empirical properties of order flow and its consequences for the modeling and understanding of market impact. 

Understanding the inter-temporal properties of the order flow is quite challenging. This is due to the intrinsic multi dimensional structure (each type of order is a different variable) whose properties depend on the current state of the book, which in turn is the result of the past order flow. The order flow and limit order book modeling is still an open problem and new models and empirical results are continuously proposed. 

Here we focus on a small but important subset of the order flow. Specifically we shall consider the flow of market orders (hence the orders triggering a transaction) and we discard the information on the volume of the order, focusing only on its sign. The sign $\epsilon(t)$ of the $t$-th trade is equal to $+1$ for a buy order and $-1$ for a sell order. Thus the binary time series of $+1$ and $-1$ is the simplest encoding of the dynamics of supply and demand for a given asset. 
One important question is what are the time autocorrelation properties of such time series. Technically speaking the autocorrelation function of the signs $\epsilon(t)$ is the expectation (or the mean) of the product of two transaction signs separated by $\tau$ transactions. For a totally random sequence (for example the one obtained by tossing a coin), this function is equal to zero for all $\tau$s because the occurrence of a head now does not give any information about the likelihood that a head or a tail will occur $\tau$ tosses in the future. The autocorrelation function is therefore related to the predictability of trade signs.

A series of papers in the last ten years have shown \cite{key-17,key-19,key-12} that in the vast majority of financial markets the autocorrelation function of the order flow is a slowly decaying function of the lag $\tau$, well described by a form $\tau^{-\gamma}$, where $\gamma$ is an exponent empirically found to be between $0$ and $1$.  These kind of processes are called long memory processes because it can be shown that they lack a typical time scale beyond which the process appears to be random. In other words the present state of the system is determined by the whole past history of the process. The exponent $\gamma$ measures the memory of the process, which is slowly decaying for large $\tau$.
It is possible to quantify this slowness from the empirical order flow auto-covariance: the smaller the exponents, the 
slower the decay. This slowness  is
commonly interpreted and quantified through the
Hurst exponent $H$. In the case of random diffusion, where
no memory is present and order signs are drawn randomly with equal
probability at each time-step, the Hurst exponent is $1/2$. For long-memory
processes $H$ is larger than $1/2$, as it is for the the order flow so that the it can be regarded as a super-diffusive
process, where fluctuations grows with time through an exponent
larger than $1/2$. 

The result is very robust and it has been checked independently by
different research groups and with different statistical methods.
The observed values of $H$ vary across markets but they alway
remain larger than $1/2$. Long-memory processes have been observed also in the dynamic behavior of other financial variables:
volatility of prices and trading volume have been recognized as long-memory
processes \cite{key-8}.

The observation of long memory of order flow raises two interesting questions:
(i) what is the behavioral origin for the sign persistence in the
order flow and (ii) how is it possible to reconcile the predictable order flows with market
efficiency (i.e. unpredictability of price changes).

The present explanations for long-memory fall into two classes:
the first is that the long memory of the order flow holds for each investor
and it links the persistence in the order flow with the presence of
large meta-orders that are splitted in subsequent small orders of the
same sign; the second class of explanations calls into question collective
behaviors of investors imitating each other \cite{key-29}. 
Evidence gathered so far seems to favor the first class of explanations,
in particular data show that large investors do split and execute
their orders. Furthermore predictions concerning the relation between
trade volumes, market impact and order flow are in agreement with
the first type of explanations \cite{key-1}. 

The presence of meta-orders is indeed a simple and clear explanation
for long-memory in the order book. Let us consider an investor that
has to execute a large trade and who does not want to influence the
price abruptly. Instead of placing directly a big market order that
would penetrate the order book and reach limit order at strongly unfavorable prices,
the investor prefers to split the meta-order and both gain the chance
not to influence the market and to get better prices if on average
other investors are not following the same strategy.

The other important question is how the long memory of order flow is consistent with EMH. Since a buy (sell) trade moves the price upward (downward) a persistent order flow sign time series would induce a persistent price change time series. This means that by observing that recent past price changes were, for example, typically positive one could predict that in the near future the price will move up. This (hypothetical) predictability would allow to make easy profits and is inconsistent with EMH and with empirical observation. 

A possible solution of this apparent paradox is the asymmetric liquidity mechanism  \cite{key-1}. According to it the price impact of a trade depends on the predictability of its sign. For example, if past trades were typically buys, implying that the next trade is more likely a buy than a sell, the asymmetric liquidity postulates that if the more predictable event (the buy) actually occurs, it will impact the price less than if the less likely event (the sell) occurs. In other words the price change is not fixed, but it is history dependent and adapts itself to the level of predictability of its source (the order flow). 

The asymmetric liquidity mechanism has been verified empirically \cite{key-1} and its microscopic origin has been explored and elucidated in  \cite{key-6}. In this last paper it has been shown that agents executing incrementally a large order by splitting it in a large number of trades adjust the size of these trades in order to minimize their impact on price. This adjustment becomes stronger and stronger as the execution proceeds. In other words investors decide their strategy exactly because they are conscious of their impact on price. Finally, it has been shown that this mechanism is able to reconcile the long memory of the order flow with the uncorrelated price change time series. 

In conclusion, the splitting of metaorders and its detection from the rest of the market is critical to understand the dynamics of price in financial markets. From a strategic point of view, the origin of splitting has been explained and motivated theoretically in the seminal work of Kyle \cite{kyle}: the optimal strategy for an investor with private information about the future price of an asset is to trade incrementally through time. This strategy allows earlier executions to be made at better prices and minimizes execution cost. Moreover this strategy minimizes the information leakage, i.e. it allows the trader to hide her intention (and information). This strategy can be seen as a different form of reflexivity. In fact the purpose of splitting is to modify, with the action of trading and the impact it generates, the beliefs of the other market participants. Differently from the impact case described in the previous section, here this form of interaction between action and beliefs creates a negative feedback. In the next two Sections we discuss in more detail the relations between these empirical facts and the theory of reflexivity. 

\section{Theory of social reflexivity}

In social systems agents do not merely observe but also actively participate in the
system themselves, this is the simple observation leading to reflexivity theory in social sciences. Soros first exposed his theory in his book \textit{The Alchemy of Finance} in 1987 where he builds his conceptual framework on two  principles. The first one is the principle of fallibility: in social systems the participants' views and consequently their perspectives are always biased, inconsistent, or both. 

Human beings are utterly familiar with the principle of fallibility that affects all aspects of human life, from perception to action. Fallibility is strictly connected to the world {\it complexity}; social facts are so complex that cannot be perceived and understood completely and thus they require a simplification that introduces biases and inconsistencies in our understanding. But fallibility is not just about perception. 

If we interpret fallibility as \textit{the inability of a human being to elaborate an optimal response in a given social situation} then we can distinguish different causes of fallibility:  
(i) A subjective cognition fallibility: we are unable to act in the best possible way because we perceive our situation in a subjective and incomplete fashion; (ii) a general cognition fallibility: we perceive the situation of all the other social agents in a subjective and incomplete fashion; (iii) a manipulative fallibility: even assuming a perfect perception of the system situation, we can act in a wrong way. 

The second is the principle of reflexivity:  the imperfect views of participants can influence the situation to
which they relate through their actions. In particular Soros gives \cite{key-0} this specific example: "if investors believe
that markets are efficient then that belief will change the way they invest, which in turn
will change the nature of the markets in which they are participating (though not
necessarily making them more efficient)." 

The role of reflexivity can be described with a circle of social actions, all affected by fallibility: the social agent perceives a given situation, formulates an interpretation, decides a strategy, and finally acts but by acting she changes the situation so that the act might no longer have the same effect that was hypothesized at first. 

This circle can generate two kinds of feedback loops, negative or positive. Negative feedback loops of participants' actions can stabilize the system, in the sense that the situation of the system becomes more and more similar to the participants' perceptions thanks to their actions, thus helping to reach an equilibrium. 

Conversely positive feedback loops destabilize the system, since participants' perceptions and the real situation of the system differ bringing the system \textit{far from equilibrium} towards an eventual sudden change, e.g. the case of \textit{bubbles} and \textit{crashes} in financial markets. 

Indeed financial markets are excellent and profitable laboratories to test reflexivity theory, and cases of positive feedback loops are investigated in \cite{key-13}. 
In the next section we will discuss some of the previously described properties of the price formation mechanism, market impact in the order book and order splitting, as examples of reflexivity where positive and negative feedback loops might play a major role. 

\section{Discussion on reflexivity and price formation}

In the first two sections we have given a general introduction and
a specific description of the intriguing phenomenology of price formation in financial markets, limit order books, market impact, and order flow. We
have discussed the possible origin of market impact, and the still open issue of its cause. We have also presented the properties of order flow, i.e. the dynamics of supply and demand arriving into the market. We have presented the long-memory property of order flow and explained it as a consequence of the splitting of meta-orders by investors. Furthermore
we have shown how the dependence of market impact from trade predictability
can explain the coexistence of long-memory in the order flow and the
fast decay of autocorrelation of price changes. Even if the description is by necessity short and synthetic, we hope that we convinced the readers that the process of price formation is interesting and still not fully understood. It is obviously at the heart of many economic (and not necessarily only financial) phenomena and has a significant number of practical consequences. 

In the text we have also proposed some analogies between the some elements of price formation and the theory of social reflexivity (reviewed in Section 3). The decisions of investors in financial markets depend on their beliefs on the traded assets and clearly price plays an important role of signal in this cognitive activity of agents. However price is affected, via market impact, by the decision of the investors and this creates the simultaneous presence of the manipulative and the cognitive function of humans, a key condition, according to Soros, for social reflexivity \cite{key-0,key-1}. We therefore believe that (financial) market microstructure is a perfect playground for studying reflexivity and understanding feedback loops between these two functions. 

More in detail, in this chapter we have sketched two possible mechanisms for reflexivity in price formation. The first is at the core of microstructure, since it concerns the origin of market impact. In fact, price moves in response to informed trades, but, at least on the short term, it moves also mechanically as a consequence of trading orders. Since other market participants cannot discern informed trades, also uninformed trades are able to move the price. This manipulative function modifies the cognitive activity of the other market participants, who revise their valuation of the asset as a consequence of the impact of a (possibly uninformed) trade. This process creates a positive feedback loop where small price fluctuations, even when generated by uninformed trades, can be amplified by the activity of reflexive agents, and this process can create price bubbles or short term large price fluctuations.

The second mechanism for reflexivity is related to the activity of order splitting. We have seen that, in consequence of the small liquidity present in financial markets and of the existence of market impact, investors who want to trade large quantities split their orders in small pieces and trade them incrementally over periods of time ranging from few minutes to several days. Other agents continuously monitor the flow of orders arriving to the market with the objective of identifying such splitting activities. This is because (i) the splitting activity of a large investor can signal her information on the future value of the price and (ii) knowing that a large investor is unloading an order gives the opportunity of front loading the investor, i.e. trading quickly with the hope to be the counterpart of the large investor in the future (and realizing a profit). Also in this case there are several interactions between the cognitive and manipulative functions of the agents. The large investor has a belief on the future price of an asset and through her trading moves the price in that direction. The other agents monitor the price and the order flow, learning the presence of a large investor and her belief on the price. Through their trading activity they modify the price pattern which in turn can modify the beliefs of other agents (and even of the splitting strategy of the large trader).

The two mechanisms presented here are clearly not exhaustive of the possible role of reflexivity in price formation and market microstructure. One of the lessons that can be learnt from this type of analysis is that the knowledge and modeling of the detailed mechanism through which agents interact is critical to understand some of the most important processes in economics and interaction of social agents. The second is that quantitative analysis of data is fundamental to validate qualitative hypothesis on investors' behavior in the market, to connect these micro-structural behaviors to market efficiency, and to formulate new hypotheses about the founding features of social systems.

\end{document}